\documentclass[iop]{emulateapj}

\usepackage{amsmath,amssymb,bm}
\usepackage[colorlinks]{hyperref}
\usepackage[all]{hypcap}
\usepackage{graphicx,float,epsfig}
\usepackage{mathrsfs,wasysym}
\usepackage{epstopdf}
\usepackage{comment}
\usepackage{pbox}
\usepackage{array}
\usepackage{xspace}
\usepackage[usenames,dvipsnames]{color}

\DeclareMathAlphabet\mathbfcal{OMS}{cmsy}{b}{n}

\DeclareSymbolFontAlphabet{\mathrsfs}{rsfs}
\DeclareMathAlphabet{\mathcal}{OMS}{cmsy}{m}{n}

\DeclareSymbolFont{bbold}{U}{bbold}{m}{n}
\DeclareSymbolFontAlphabet{\mathbbold}{bbold}

\newcommand{\be}{\begin{equation}}
\newcommand{\ee}{\end{equation}}

\newcommand{\pd}{\partial}

\def\go{\mathrel{\raise.3ex\hbox{$>$}\mkern-14mu
             \lower0.6ex\hbox{$\sim$}}}
\def\lo{\mathrel{\raise.3ex\hbox{$<$}\mkern-14mu
             \lower0.6ex\hbox{$\sim$}}}

\newcommand{\tline}{\texorpdfstring{\\}{ }}

\graphicspath{{figs/}}

\begin{document}


\title{``Slimplectic'' Integrators: variational integrators\tline{}for general nonconservative systems}

\author{David Tsang \altaffilmark{1,$\dagger$}, Chad R.\!~Galley\altaffilmark{2}, Leo C. Stein\altaffilmark{3,4}, Alec Turner\altaffilmark{1}}

\altaffiltext{1}{Department of Physics, McGill University, Montreal, QC, H3A 2T8, Canada}
\altaffiltext{$\dagger$}{\href{mailto:dtsang@physics.mcgill.ca}{dtsang@physics.mcgill.ca}}
\altaffiltext{2}{Theoretical Astrophysics, Walter Burke Institute for Theoretical Physics, California Institute of Technology, Pasadena, CA 91125, USA}
\altaffiltext{3}{Center for Radiophysics and Space Research, Cornell University, Ithaca, NY 14853, USA}
\altaffiltext{4}{Einstein Fellow}

\setcounter{front@matter@foot@note}{-1}

\begin{abstract}
  Symplectic integrators are widely used for long-term integration 
  of conservative astrophysical problems due to 
  their ability to preserve the constants of motion; however, they cannot in general 
  be applied in the presence of nonconservative interactions. 
  In this Letter, we develop the ``slimplectic'' integrator, a new type of numerical
  integrator that shares many of the benefits of traditional
  symplectic integrators yet is applicable to general
  nonconservative systems. We utilize a fixed-time-step variational integrator
  formalism applied to the principle of stationary nonconservative
  action developed in~\citet*{Galley2013, Galley2014}. As a result, the 
  generalized momenta and energy (Noether current) evolutions are well-tracked. 
  We discuss several example systems, including damped
  harmonic oscillators, Poynting-Robertson drag, and gravitational radiation reaction, 
  by utilizing our new publicly available code to
  demonstrate the slimplectic integrator algorithm.

  Slimplectic integrators are well-suited for integrations of systems
  where nonconservative effects play an important role in the
  \emph{long-term} dynamical evolution. As such they are particularly
  appropriate for cosmological or celestial N-body dynamics problems where nonconservative
  interactions, e.g.~gas interactions or dissipative tides, can play an important role.

\end{abstract}

\section{Introduction}
Symplectic integrators are a class of mappings that allow for
numerical integration of conservative dynamical systems and which, up to round-off, exactly
preserve certain constants of motion (e.g.~the symplectic form). As a result the integrations 
do not suffer from numerical ``dissipation'' which would cause an unphysical drift 
over many dynamical times.
 Due to these properties, symplectic integrators
 are widely used in the long-term integration of many physical systems, 
 particularly in celestial dynamics~\citep{Wisdom1991, Gladman1991, Levinson1994, Tamayo2015}.  
 
Conservative variational integrators~\citep[see e.g.][]{Marsden2001}
are a subclass of symplectic integrators where the mappings are
determined by the extremization of a \emph{discretized} action.\footnote{%
  Most symplectic integrators can be written as (local) variational
  integrators.
}
The discretized action can inherit the symmetries of the full action
such that, by Noether's theorem, the discrete equations of motion exactly conserve the
symplectic form and the momenta. Since discretizing the time coordinate 
breaks the continuous time-shift symmetry, fixed-time-step
variational integrators that preserve the symplectic form and the momenta cannot
also conserve energy~\citep{ge1988}.  However, the energy
error tends to be bounded by a constant, even over long integration
times \citep[see e.g.][and references therein]{lew2004}, in contrast with traditional integration methods where the
error tends to grow with time.

Variational integrators can be applied to some dissipative problems
using the Lagrange-d'Alembert approach~\citep{Marsden2001, lew2004}.
Here, we utilize the more general 
nonconservative action principle, recently developed
by~\citet*{Galley2013,Galley2014}.  This
formalism was developed to accommodate generically the causal dynamics of untracked 
or inaccessible degrees of freedom that might result from an
integrating-out or coarse-graining procedure at the level
of an action/Lagrangian/Hamiltonian.

In this Letter, we develop variational integrators
from the nonconservative action principle. The resulting mappings are no
longer symplectic, as the symplectic form (and momenta) are no longer
conserved, but evolve according to the nonconservative dynamics.  We
instead refer this type of numerical integrator as ``slimplectic''
since phase space volumes tend to slim down for dissipative systems.
We show that our method inherits many of the
same performance features of the symplectic integrator. Previous works have demonstrated some success by including weakly dissipative forces in 2nd-order symplectic integrators \citep[see e.g.][]{Malhotra1994, Cordeiro1996, Mikkola1997, Hamilton1999, Zhang2007}; in fact, these can be shown to be particular cases of our more general (arbitrary order) method. 
For brevity, we focus 
on a basic fixed-time-step slimplectic integrator leaving further
developments, such as  adaptive time-stepping and detailed discussion
of Noether current evolution, to a longer follow-up paper.

\section{Nonconservative Lagrangian Mechanics}
Nonconservative Lagrangian mechanics accommodates nonconservative interactions 
and effects by first formally doubling the degrees of freedom, $q \rightarrow (q_1, q_2)$.
The action describing the dynamics of these doubled variables is 
\begin{align}
{\cal S} = \int \Lambda(q_{1,2}, \dot{q}_{1,2}, t) dt
\end{align}
where the (nonconservative) Lagrangian is
\begin{align}
\Lambda(q_{1,2}, \dot{q}_{1,2}, t) \! = \! L(q_1, \dot{q}_1, t) \!-\! L(q_2, \dot{q}_2, t) \!+\! K(q_{1,2}, \dot{q}_{1,2}, t). \nonumber
\end{align}
$L$ is the usual Lagrangian, which is an arbitrary function of coordinates, 
velocity, and time and describes the conservative sector of the system 
(i.e., dynamics in the absence of nonconservative effects). 
However, $K$ is another arbitrary function that couples the variables together, 
vanishes when $q_1 = q_2$, and accounts for any generic nonconservative
interaction. Note that $\Lambda$ is completely specified once $L$ and $K$
are given.

After all variations of ${\cal S}$ are performed the two variables are identified
with each other, $q_1 = q_2 = q$, which is called the \emph{physical limit} (PL). 
In some cases, it is convenient to work with more physically motivated
coordinates, $q_- = q_1-q_2$ and $q_+ = (q_1+q_2)/2$. The former
can often be considered like a virtual displacement and vanishes in the PL while the latter is
the surviving physically relevant combination.

Requiring ${\cal S}$ to be stationary under variations of the doubled variables
and taking the PL leads to nonconservative Euler-Lagrange equations of motion,
\begin{align}
 \left[\frac{d}{dt} \frac{\pd \Lambda}{\pd \dot{q}_-} - \frac{\pd \Lambda}{\pd q_-} \right]_{\rm PL} &= 0 
\intertext{or in terms of $L$ and $K$, }
\frac{d}{dt} \frac{\pd L}{\pd \dot{q}} - \frac{\pd L}{\pd q} &= \left[ \frac{\pd K}{\pd q_-}  - \frac{d}{dt} \frac{\pd K}{\pd \dot{q}_-} \right]_{\rm PL}  .
\end{align}
There are multiple ways to specify $K$, which depend on the particular problem in question. 
More details about this aspect and the nonconservative action
formalism in general can be found in {\S}II of~\citet*{Galley2014}, including the 
evolution of Noether currents according to nonconservative
processes described by  $K$.

\section{Variational Integrators for Nonconservative Systems}

Variational integrators numerically approximate the behavior of a
system by implementing the exact equations of motion for a
closely-related \emph{discrete} action~\citep{Marsden2001, Brown2006}.  
Here we will apply it to the nonconservative action
principle described above.

To construct a variational integrator we need to make a choice of
discretization of the action integral
${\cal S} = \int_{t_i}^{t_f} \Lambda(q_\pm, \dot{q}_\pm, t) dt$. 

\begin{figure}
  \includegraphics[width=1.0\columnwidth]{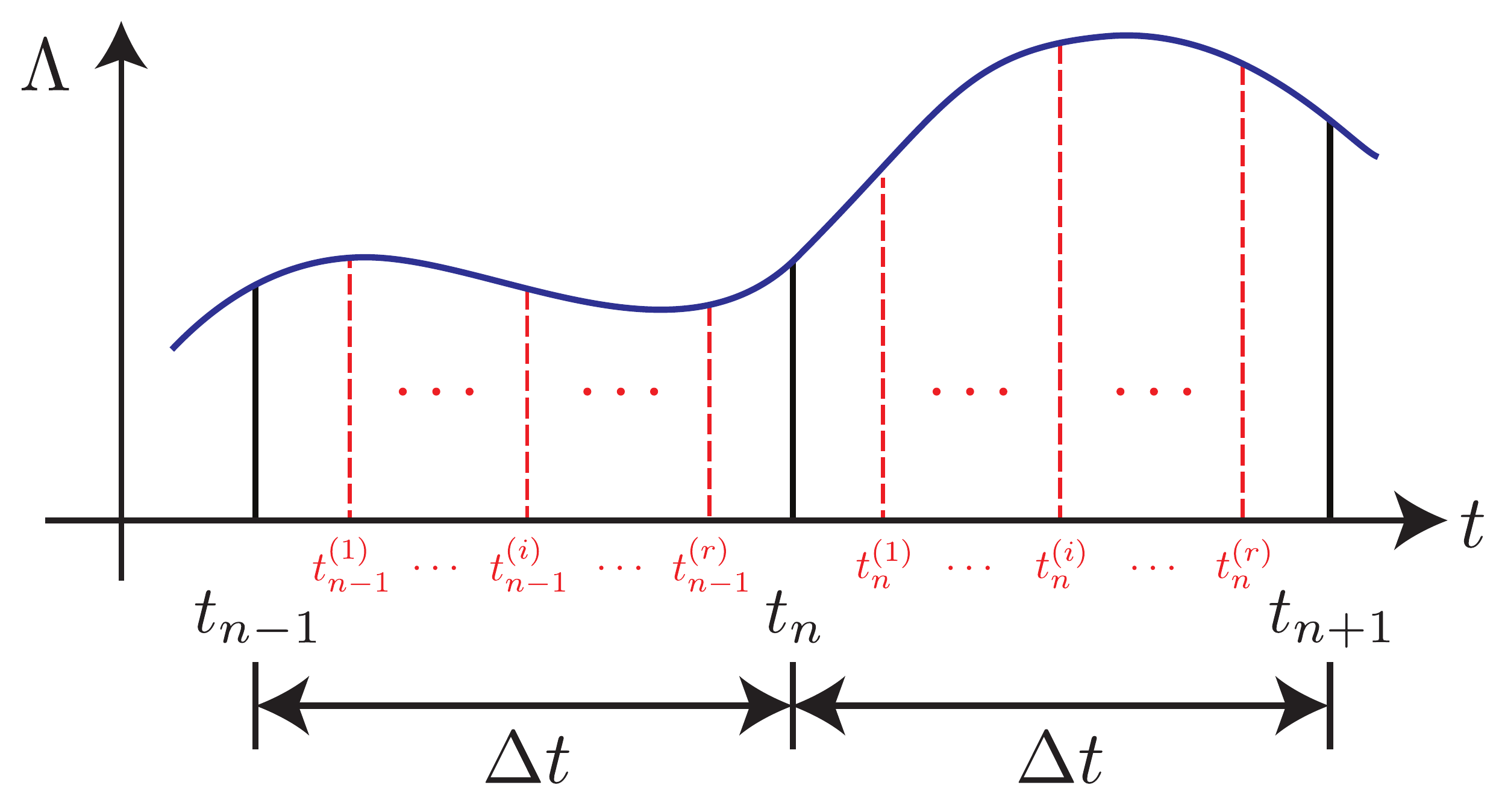}
  \caption{The time discretization used for the Galerkin-Gauss-Lobatto (GGL)
    variational integrator. The action is approximated using the
    Gauss-Lobatto quadrature method, where for each interval
    $\Delta t$, the area under the nonconservative Lagrangian is approximated by the weighted
    sum of the Lagrangian, $\Lambda(q_\pm, \dot{q}_\pm, t)$, evaluated at each of
    the $(r + 2)$ quadrature points. This quadrature rule (and the variational integrator) is accurate
    up to order $(2r + 2)$. 
    The generalized velocities $\dot{q}_\pm$ are approximated using the derivative of
    the $(r + 1)$th-order interpolating function at each quadrature point. 
    }
  \label{fig:GGLTimesLambda}
\end{figure}

A choice that provides a time-reversal symmetric
discretization~\citep[thus an even order method;][]
{Farr2007} is Gauss-Lobatto quadrature, illustrated in
Figure~\ref{fig:GGLTimesLambda}.  On the time interval $t\in
[t_{n},t_{n+1}]$ with $\Delta t = t_{n+1}-t_{n}$, we
have the set of $r + 2$ quadrature points
$t_n, \{t_n^{(i)}\}_{i=1}^r, t_{n+1}$, with
\begin{align}
t_n^{(i)} \equiv t_n + (1 + x_i)\frac{\Delta t}{2},
\end{align}
where $x_0 \equiv -1$, $x_{r+1} \equiv +1$, and $x_i$ (for
$i \in \{1 \ldots r\}$) is the $i$th root of $dP_{r+1}/dx$, the
derivative of the $(r+1)$th Legendre polynomial, $P_{r+1}(x)$.  For a
given nonconservative Lagrangian functional
$\Lambda(q_\pm, \dot{q}_\pm, t)$, we can approximate the degrees of
freedom $q_{n,\pm}(t) = \phi_{n,\pm}(t) + {\cal O}(\Delta t^{r+2})$
using the cardinal-function interpolation for this choice of
quadrature points.

We then have the approximation
$\dot{q}_{n,\pm}(t) \simeq \dot{\phi}_{n, \pm}(t)$, which can be
conveniently evaluated at the quadrature points using the derivative
matrix~\citep[see e.g.][]{Boyd2001,BoydErrata},
\begin{align}
D_{ij} = \begin{cases}
  \hfil -{(r+1)(r+2)}/({2\Delta t})&\qquad i = j = 0 \\
  \hfil {(r+1)(r+2)}/({2\Delta t}) &\qquad i = j = r+1 \\
  \hfil 0 &\qquad i= j \neq 0, r+1\\
  \hfil \displaystyle\frac{2P_{r+1}(x_i)}{P_{r+1}(x_j)(x_i - x_j) \Delta t} &\qquad i \neq j
  \end{cases}
\end{align}
such that
\begin{align}
\dot{\phi}_{n,\pm}(t_n^{(i)}) = \sum_{j=0}^{r+1} D_{ij} \,q_{n,\pm}^{(j)} \equiv \dot{\phi}_{n,\pm}^{(i)}
\end{align}
where for notational compactness we define $t_n^{(0)} \equiv t_n$,
$t_n^{(r+1)} \equiv t_{n+1}$, and
$q_{n,\pm}^{(i)} \equiv q_\pm(t_n^{(i)})$.

Using Gauss-Lobatto quadrature, any integral functional $\int F dt$
[for example $F\in \{\Lambda, L, K\}$] has a discrete-quadrature
approximation on the time interval $[t_{n},t_{n+1}]$.  This
discrete functional $F_d^n$ is\footnote{%
For $L_d^n$ we can drop the $\pm$, indices.}
\begin{align}
F_d(q_{n,\pm},  \{q_{n,\pm}^{(i)}\}_{i=1}^r, q_{n+1,\pm}, t_n) &\equiv \sum_{i=0}^{r+1} w_i F(q_{n,\pm}^{(i)}, \dot{\phi}_{n,\pm}^{(i)}, t_n^{(i)}) \nonumber\\
 &\equiv F_d^n,
\end{align}
where the Gauss-Lobatto quadrature weights $w_i$ are given by
\begin{align}
w_i \equiv \frac{\Delta t}{(r+1)(r+2)[P_{r+1}(x_i)]^2}.
\end{align}

Now we approximate the action over an interval $[t_0, t_{N+1}]$,
\begin{align}
{\cal S}[t_0, t_{N+1}] &= \int_{t_0}^{t_{N+1}} \Lambda(q_\pm, \dot{q}_\pm, t) dt \nonumber \\
&= {\cal S}_d[t_0, t_{N+1}]  + {\cal O}(\Delta t^{2r+3}),
\end{align}
where the discretized action is defined as
\begin{align}
{\cal S}_d[t_0, t_{N+1}]  \equiv \sum_{n=0}^{N} \Lambda_d(q_{n, \pm}, \{q_{n,\pm}^{(i)}\}_{i=1}^r, q_{n+1,\pm}, t_n)
\label{eq:DiscreteAction}.
\end{align}
We refer to this discretization choice as the Galerkin-Gauss-Lobatto (GGL) method~\citep{Farr2007}. 

The discrete action ${\cal S}_d[t_0, t_{N+1}]$ from~\eqref{eq:DiscreteAction} 
can then be extremized over values
$q_{n,-}$ and $q_{n, -}^{(i)}$, and the physical limit imposed, to
generate the discretized equations of motion for each $n \in [1, N]$,
\begin{subequations}
\begin{align}
\left[\frac{\pd  \Lambda_d^{n-1}}{\pd q_{n,-}} + \frac{\pd \Lambda_d^n}{\pd q_{n,-}}\right]_{\rm PL} &= 0, \\
\left[\frac{\pd \Lambda_d^n}{\pd q_{n,-}^{(i)}}\right]_{\rm PL} &= 0,
\end{align}
\end{subequations}
since from Figure~\ref{fig:GGLTimesLambda}, we see that each
$q_{n,-}^{(i)}$ only contributes to a single $\Lambda_{d}^{n}$ in the discretized
action~\eqref{eq:DiscreteAction}, while each $q_{n,-}$ appears in both 
$\Lambda_{d}^{n-1}$ and $\Lambda_{d}^{n}$.

In terms of $L_d^n$ and $K_d^n$, the equations of motion are 
\begin{subequations}
\label{eq:discreteEOMs}
\begin{align}
\frac{\pd  L_d^{n-1}}{\pd q_n} + \frac{\pd L_d^n}{\pd q_n}  + \left[\frac{\pd  K_d^{n-1}}{\pd q_{n,-}} + \frac{\pd K_d^n}{\pd q_{n,-}}\right]_{\rm PL} &= 0 \label{eq:FullDiscreteNCEOM}\\
\frac{\pd L_d^n}{\pd q_n^{(i)}}  + \left[\frac{\pd K_d^n}{\pd q_{n,-}^{(i)}}\right]_{\rm PL} &= 0.
\end{align}
\end{subequations}%
We now introduce the discrete momenta $\pi_{n}$, defining the nonconservative (slimplectic) GGL variational integrator
map $(q_n, \pi_n) \rightarrow (q_{n+1}, \pi_{n+1})$, by splitting the equation
of motion \eqref{eq:FullDiscreteNCEOM} into
\begin{subequations}
\label{eq:Discrete_NC_EOM}
\begin{align}
\pi_{n} &\equiv - \left[ \frac{\pd \Lambda_d^n }{\pd q_{n,-}} \right]_{\rm PL} &={}& - \frac{\pd  L_d^n }{\pd q_n}- \left[ \frac{\pd K_d^n }{\pd q_{n,-}} \right]_{\rm PL} \label{eq:NCqMap},\\
\pi_{n+1} &\equiv \left[ \frac{\pd \Lambda_d^n }{\pd q_{n+1,-}} \right]_{\rm PL}  &={}& \frac{\pd L_d^n}{\pd q_{n+1}} + \left[ \frac{\pd K_d^n}{\pd q_{n+1,-}}\right]_{\rm PL} \label{eq:NCpiMap},\\
0 &= \left[ \frac{\pd \Lambda_d^n }{\pd q^{(i)}_{n,-}} \right]_{\rm PL} &={}&  \frac{\pd L_d^n}{\pd q_n^{(i)}}+ \left[\frac{\pd K_d^n}{\pd q_{n,-}^{(i)}} \right]_{\rm PL} \label{eq:NCintEq}.
\end{align}
\end{subequations}

Given initial values of $(q_n, \pi_n)$, the values of $q_{n+1}$ are determined implicitly by~\eqref{eq:NCqMap},
while the values for the $q_n^{(i)}$ intermediate points are given
implicitly by equation~\eqref{eq:NCintEq} for $i \in \{1 \ldots r\}$.
The final momenta $\pi_{n+1}$ can then be determined explicitly
from~\eqref{eq:NCpiMap}.

Noether's theorem for conservative actions can be shown to 
generalize to nonconservative systems where the corresponding 
Noether currents evolve in time due to a non-zero K~\citep*{Galley2014}. 
One can show that for continuous symmetries of the
conservative action, which remain after discretization, discrete
Noether currents will also evolve due to $K_d$. 
Thus, translational or rotational symmetries, for example, will generate 
discrete momenta that evolve according to $K_d$, up to round off and bias error~\citep{Brouwer1937, Rein2015}. 
Additional error compared to the physical evolution is only due to the discretization of the action. 

\begin{figure}
\centering
\includegraphics[width=\columnwidth]{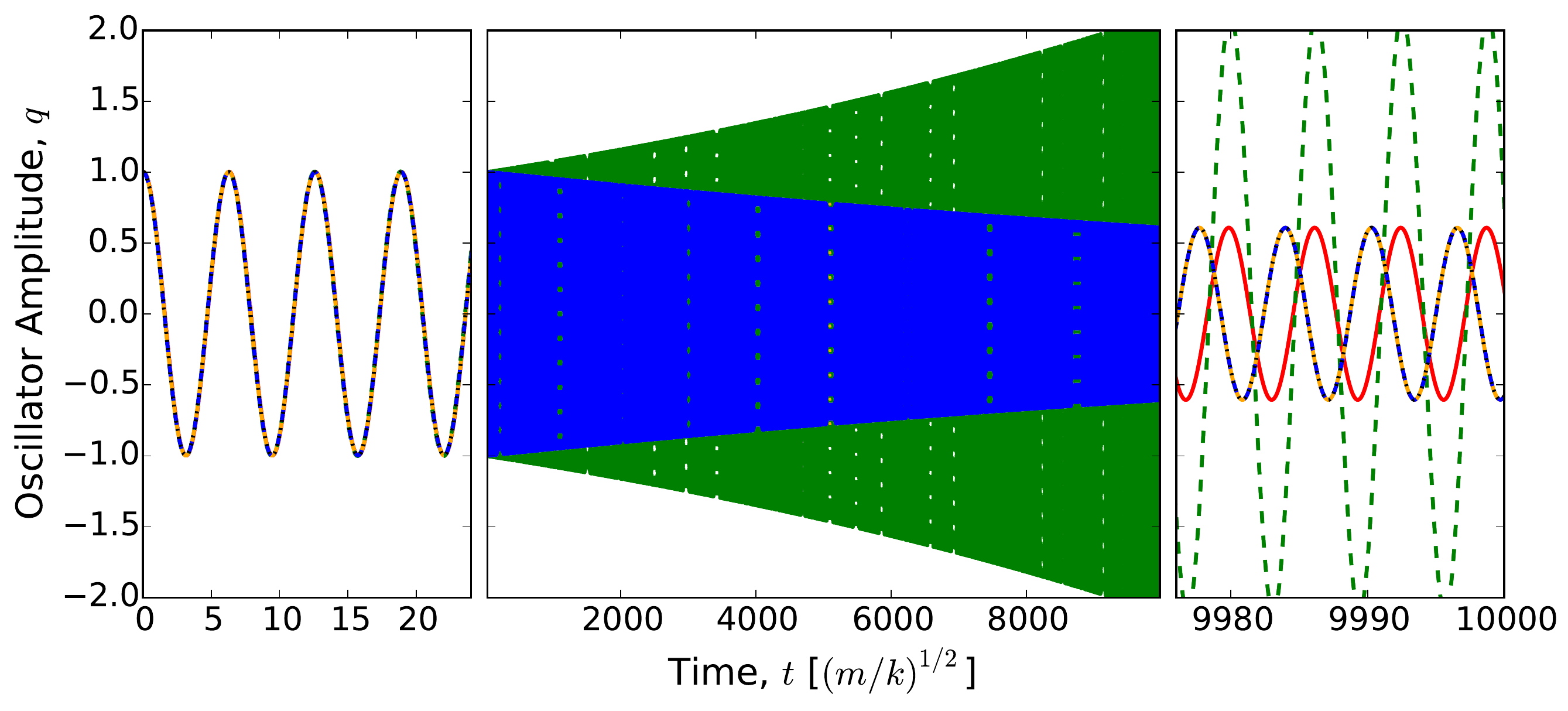}
\includegraphics[width=1.015\columnwidth]{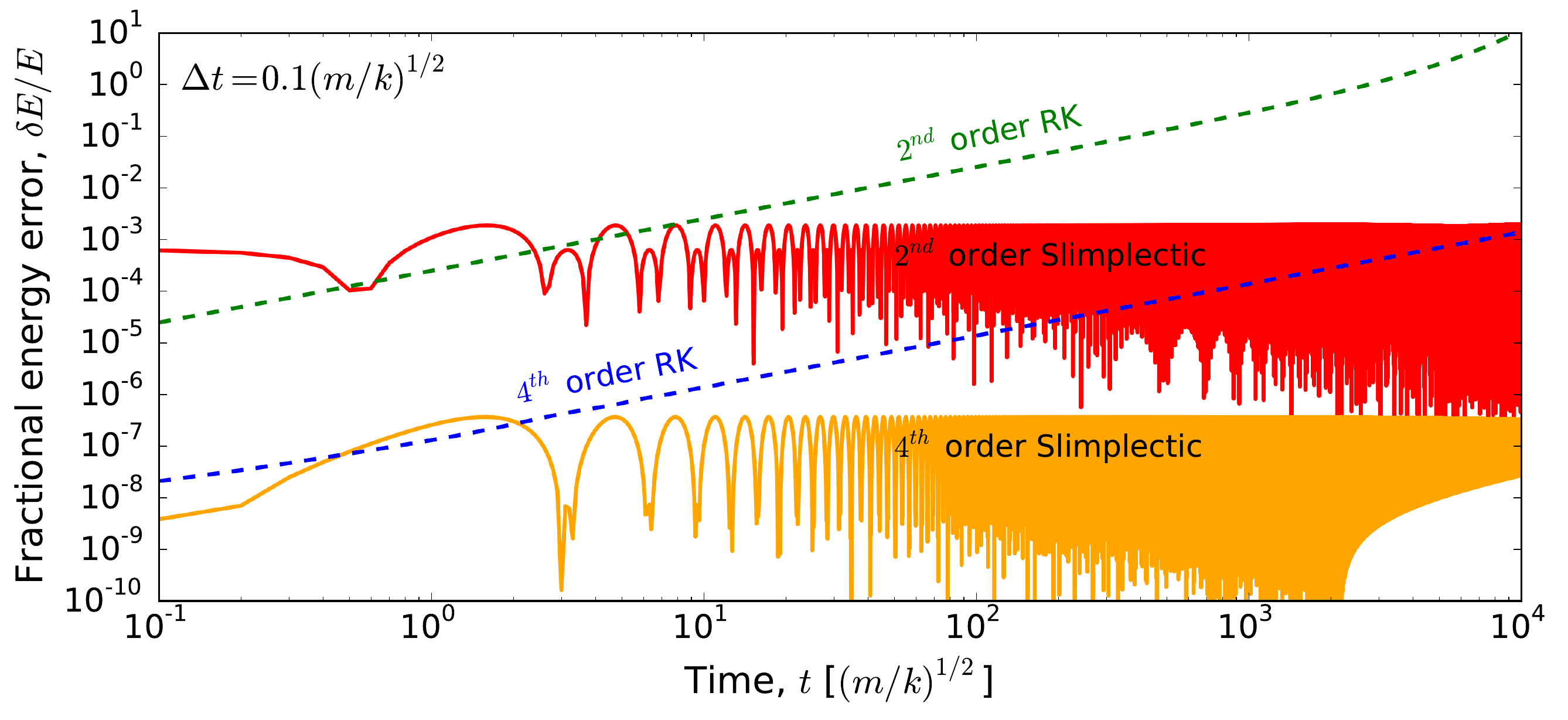}
\caption{%
{\bf Top:} Numerical solutions for the damped harmonic oscillator, 
described by $L = m \dot{q}^2/2 - k q^2/2$,
$K = - \lambda \dot{q}_+ q_-$, with $\lambda = 10^{-4} (mk)^{1/2}$,
and fixed time steps $\Delta t = 0.1 (m/k)^{1/2}$. Initial conditions
were taken to be $q(0) = 1$ and $\dot{q}(0) = 0$. The $2$nd
order RK solution, shown in green, is unstable for these parameters
and diverges significantly. The $4$th-order RK solution (blue-dashed)
cannot be readily distinguished from the $4$th-order slimplectic solution (solid-orange) in this plot. The $2$nd-order
slimplectic solution, (solid-red), gives nearly the correct amplitude
after $\sim10^5$ time steps, as the energy evolution is accurately followed, 
though a phase shift is evident. 
{\bf Bottom:} The fractional energy error relative to
the energy given by the analytic solution at each time. We see that while
initially the RK and slimplectic energy errors are comparable at each
order, the RK energy errors grow roughly linearly with time, while the
slimplectic energy error remains bounded.
\label{fig:DSHO}
}
\end{figure}

The GGL discretization does \emph{not} preserve the
time-shift symmetry preventing energy evolution from being precisely
tracked. However, the fractional energy error tends to be oscillatory
and bounded by a resolution and order-dependent constant.\footnote{%
Fixed-time step variational integrator methods
cannot be both symplectic-momentum and momentum-energy
preserving~\citep{ge1988}, however adaptive time-stepping allows
symplectic-energy-momentum methods to be
developed~\citep{kane1999,Preto1999, lew2003}.
}
We will defer more detailed discussion of Noether current evolution to
a longer followup paper in the interests of space.

The resulting slimplectic maps are accurate up to order $2r + 2$. For
$r = 0$, where no intermediate steps are used, the quadrature method
is the trapezoid rule, and the variational integrator is 2nd-order and
equivalent to the St\"ormer-Verlet ``leap-frog''
integrator~\citep{Wendlandt1997}.

It is well known that 2nd-order ``leap-frog'' integrators can be used
for dissipative systems, by inserting a dissipative ``kick'' force
into the ``kick-drift-kick'' ansatz, resulting in good energy and
momentum evolution properties. Our approach explains \emph{why} this
simple modification works in the 2nd-order system, as it is equivalent
to the lowest order slimplectic GGL method \citep[see also][for a similar Lagrange-d'Alembert approach]{lew2004}. The slimplectic method allows this to be
generalized to higher orders and general nonconservative systems\footnote{Similar results can be obtained for Wisdom-Holman-type mappings by splitting the nonconservative action into integrable and perturbative  terms \citep[see e.g.][]{Farr2009}.}.


\section{Code and Examples}
We have developed a simple \texttt{python} code,
\texttt{slimplectic}, that is publicly available\footnote{%
\label{fn:repo}%
The repository for \texttt{slimplectic} is available at
\url{http://github.com/davtsang/slimplectic}.
}
and generates the fixed-time-step
slimplectic GGL integrators described above, for use in characterizing
the numerical technique. The code generates slimplectic solvers of arbitrary order
$(2r + 2)$ given \texttt{sympy}~\citep{sympy} expressions for
$L(q, \dot{q}, t)$ and $K(q_\pm, \dot{q}_\pm, t)$.  This demonstration
code is designed to work for arbitrary $L$ and $K$, and thus has not
been optimized as would be appropriate for specific problems.  In particular, the
equations of motion~\eqref{eq:Discrete_NC_EOM} are solved with
standard root-finders, rather than a problem-specific iteration
method, to be more generally
applicable.  

As a basic example in Figure~\ref{fig:DSHO} we compare
Runge-Kutta (RK) and slimplectic integration of a simple damped harmonic
oscillator, for both 2nd and 4th-order methods.  Below we also present
two basic astrophysical examples of non-conservative interactions. 
All examples are available as \texttt{ipython} notebooks in our public 
repository.\textsuperscript{\ref{fn:repo}}

\subsection{Poynting-Robertson Drag}

\begin{figure}
\centering
\includegraphics[width=\columnwidth]{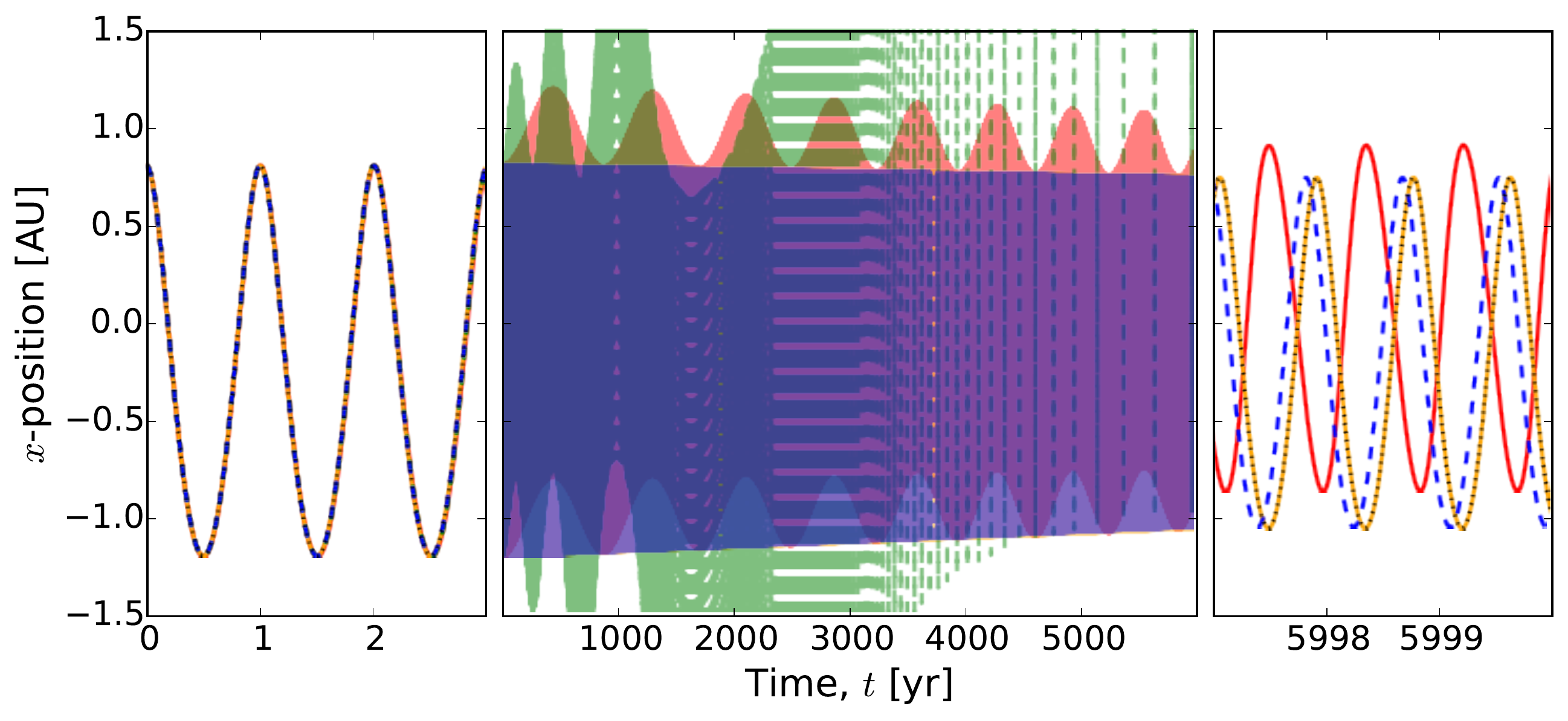}
\includegraphics[width=\columnwidth]{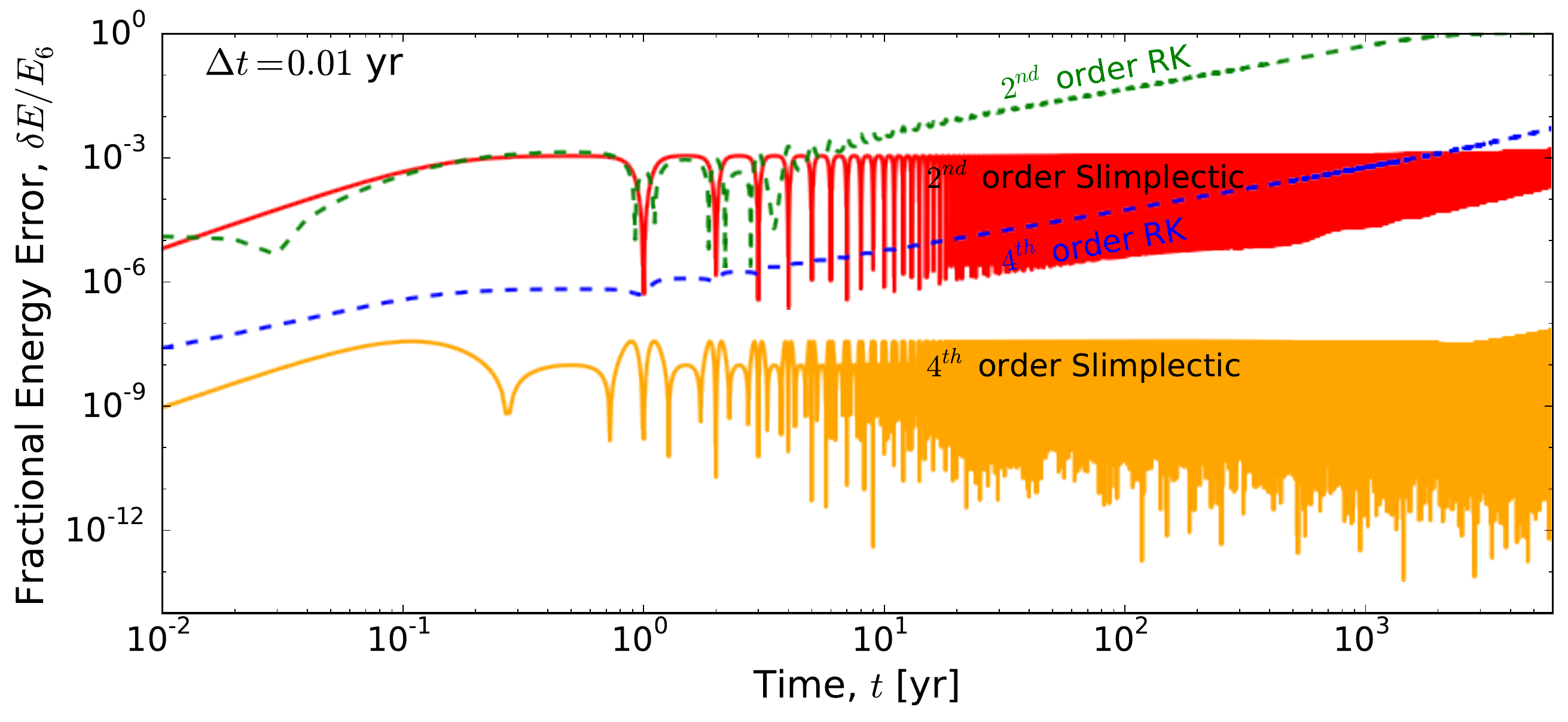}
\caption{%
{\bf Top:} Evolution of the Cartesian x-coordinate for orbital motion
of a particle experiencing Poynting-Robertson drag due to radiation from a solar-type star,
 with particle density $\rho = 2$ g cm$^{-3}$ and particle size
$d = 5\times 10^{-2}$ cm. The particle has initial semi-major axis of
1AU, and initial eccentricity of $0.2$. With a fixed-time-step of
$\Delta t = 0.01$ yr, the 2nd-order RK integrator (green-dashed) is
unstable, while the 2nd-order slimplectic integrator (red) behaves
much more accurately, despite significant (numerical) precession over
a $\sim$1000 year timescale. The 4th-order integrators (RK, blue-dashed;
slimplectic, orange) have similar long-term amplitude evolution
without visible numerical precession. The phase errors (not shown) of the RK
integrators grow $\propto t^2$, while the slimplectic integrators have
phase error $\propto t$ \citep[see][]{Preto1999}. 
{\bf Bottom:} Fractional energy error (compared to a 6th-order
slimplectic integration) for the system described above. The RK
errors grow roughly linearly in time,
while the slimplectic errors are bounded. There is a slight turn-up in the 4th-order slimplectic error envelope, most likely due to
the build up of round-off, bias, or action-discretization error.
\label{fig:PRDrag}
}
\end{figure}

We first examine the orbital motion of a dust particle
experiencing Poynting-Robertson drag~\citep{Burns1979} due to radiation from a solar
type star, starting with a semi-major axis of 1AU in an eccentric
($e = 0.2$) orbit.  The Lagrangian for this system is
\begin{align}
L = \frac{1}{2} m \dot{\mathbf q}^2 + (1-\beta)\frac{GM_{\odot}m}{|{\mathbf q}|},
\end{align}
where $m$ is the dust particle's mass, and ${\mathbf q}$ its
position. The dimensionless factor
\begin{align}
\beta \equiv \frac{3L_\odot}{8\pi c \rho\,G M_\odot d} \simeq 0.058 \left(\frac{\rho}{2\, {\rm g}\,{\rm cm}^{-3}} \right)^{-1} \left(\frac{d}{10^{-3} {\rm cm}} \right)^{-1},
\end{align}
is the ratio between forces due to radiation pressure and gravity, where
$L_\odot$ and $M_\odot$ are the solar luminosity and mass, $\rho$ and $d$ are the density and size of the dust grain.
The nonconservative potential which generates the correct
Poynting-Robertson drag force is (for $|\dot{\mathbf q}| \ll c$)
found as the virtual work from the known force,
\begin{align}
K = -\frac{\beta G M_\odot m }{c\, {\mathbf q}_+^2} \left[\dot{{\mathbf q}}_+ \cdot {\mathbf q}_- + \frac{1}{{\mathbf q}_+^2}(\dot{{\mathbf q}}_+ \cdot {\mathbf q}_+)({\mathbf q}_+ \cdot {\mathbf q}_-)\right].
\end{align}
Methods to determine or derive $K$ are discussed in~\citep{Galley2014}.
The first term in square brackets above gives the usual drag term,
while the second term is due to the Doppler shift caused by radial motion.

The system was integrated for $6000$ years using 2nd and 4th-order RK (green and blue dashed) and slimplectic (red and orange
solid) methods with time-steps of $\Delta t = 0.01$ yr. The results and discussion are shown in Figure~\ref{fig:PRDrag}.

\subsection{Gravitational Radiation Reaction}

\begin{figure}
\includegraphics[width=\columnwidth]{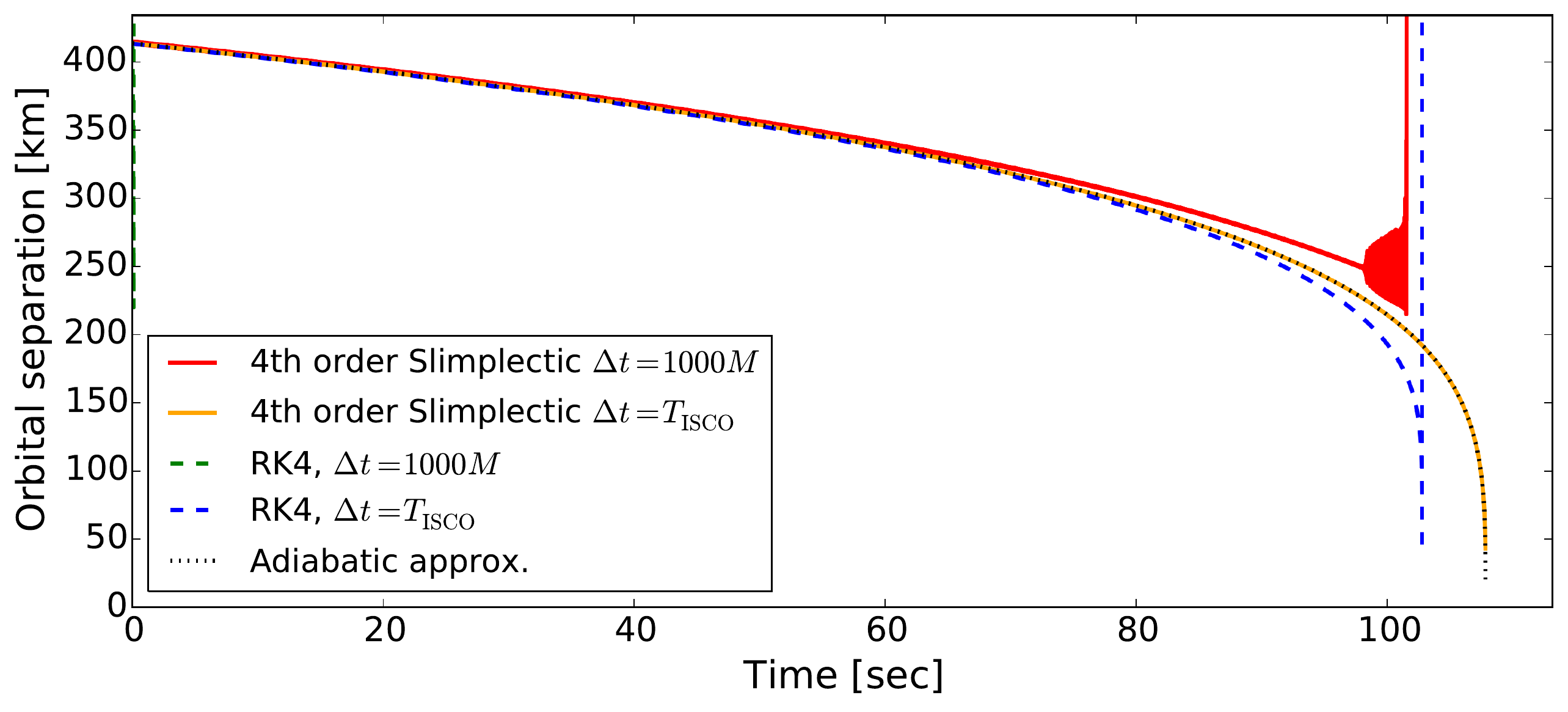}
\includegraphics[width=\columnwidth]{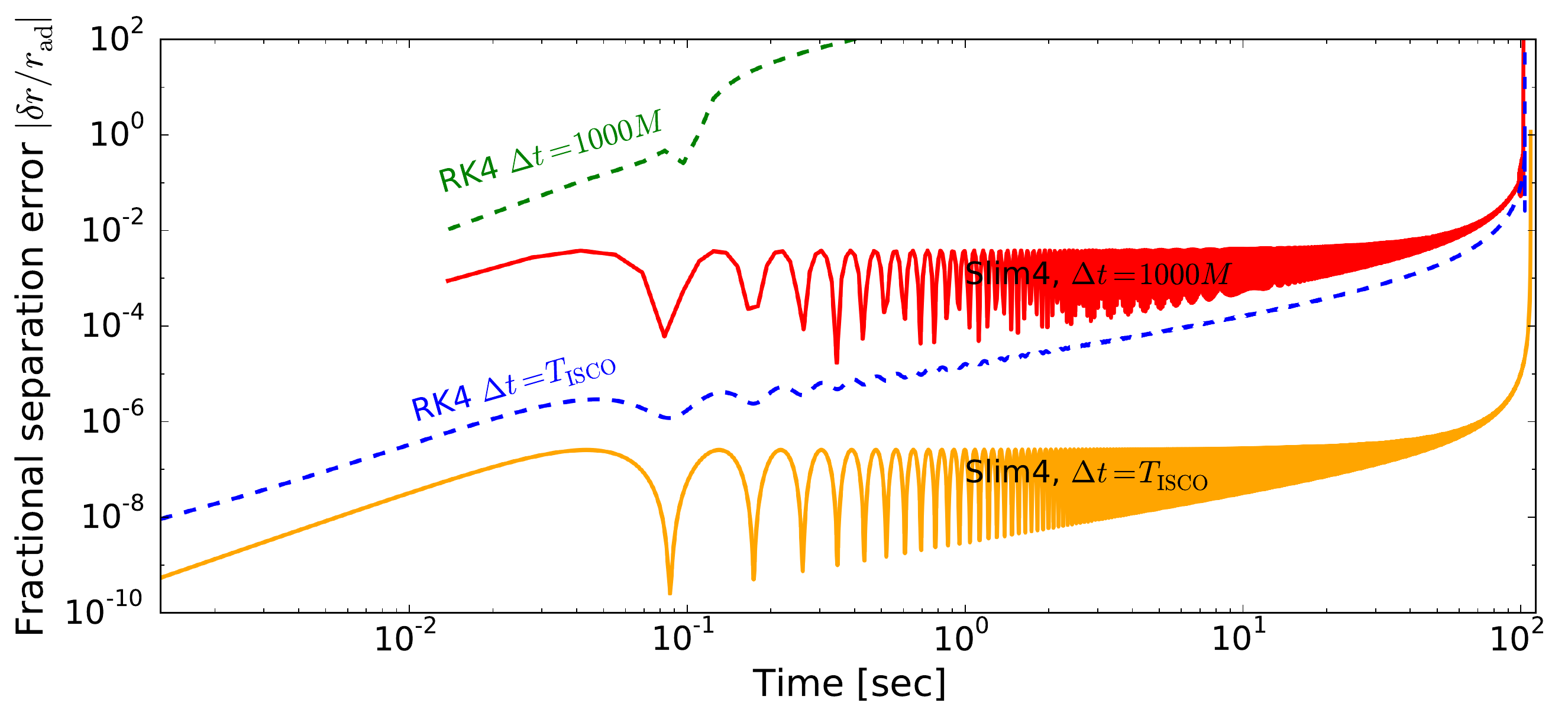}
\caption{%
{\bf Top:} PN radiation reaction, including only the 2.5PN dissipative terms
through $K$. Each of the methods (RK and slimplectic GGL)  are 4th
order with fixed timesteps of $\Delta t = 1000M$ and $\Delta t = T_{\rm
  ISCO}$. The integration methods blow up when the orbital
time-scale is comparable to the time step, though the
slimplectic method performs significantly better than the RK method
for the same time steps; for the $1000M$ time-step the
RK method is immediately unstable. \\
{\bf Bottom:} Relative error in the orbital radius compared to the analytic
adiabatic-approximation solution. \\
\label{fig:PNr}
}
\end{figure}

We also consider two $1.4 M_\odot$ neutron stars inspiraling from gravitational wave emission.
This example demonstrates (fixed-time-step) integrators for systems where
the orbital dynamics can change quickly due to nonconservative
effects. In the post-Newtonian (PN) approximation, leading-order
conservative dynamics for the orbital separation, ${\mathbf q}$, are described by Newtonian gravity
\begin{align}
L = \frac{1}{2} \mu \dot{\mathbf q}^2 + \frac{\mu M}{|{\mathbf q}|},
\end{align}
where $M=m_1 + m_2 = 2.8 M_\odot$ is the total mass,  $\mu = m_1 m_2/M
= 0.7 M_\odot$ is the reduced mass, and $G = c =1$. 

Dissipative effects from radiation reaction first appear at PN order $(|\dot{\mathbf q}|/c)^5$ (or 2.5PN) and
are described by $K$, which has been calculated in~\citet{Galley2012, Galley2009}. After order-reduction, 
\begin{align}
K ={}& \frac{16}{5}\nu^2 M^4 \frac{(\dot{\mathbf q}_+ \cdot {\mathbf q}_-)}{|{\mathbf q}_+|^4} 
- \frac{48}{5} \nu^2 M^3 \frac{|\dot{\mathbf q}_+|^2 (\dot{\mathbf q}_+ \cdot {\mathbf q}_-)}{|{\mathbf q}_+|^3} \nonumber \\
   &{}+ 24 \nu^2 M^3 \frac{(\dot{\mathbf q}_+ \cdot {\mathbf q}_+)^2 (\dot{\mathbf q}_+ \cdot {\mathbf q}_-)}{|{\mathbf q}_+|^5} \nonumber\\
   &{}+ \frac{16}{15} \nu^2 M^4 \frac{(\dot{\mathbf q}_+ \cdot {\mathbf q}_+) ({\mathbf q}_+ \cdot {\mathbf q}_-)}{|{\mathbf q}_+|^6} \nonumber\\
   &{}+ \frac{144}{5} \nu^2 M^3 \frac{|\dot{\mathbf q}_+|^2 (\dot{\mathbf q}_+ \cdot {\mathbf q}_+) ({\mathbf q}_+ \cdot {\mathbf q}_-)}{|{\mathbf q}_+|^5} \nonumber\\
   &{}- 40 \nu^2 M^3 \frac{(\dot{\mathbf q}_+ \cdot {\mathbf q}_+)^3 ({\mathbf q}_+ \cdot {\mathbf q}_-)}{|{\mathbf q}_+|^7}
\,,
\end{align}
where $\nu = \mu/M = 1/4$.
A physically consistent simulation should go to the same PN order in
$L$ and $K$.  Here, instead, we use the leading order in each as a toy
model to focus on the numerical method.

\begin{figure}
\includegraphics[width=\columnwidth]{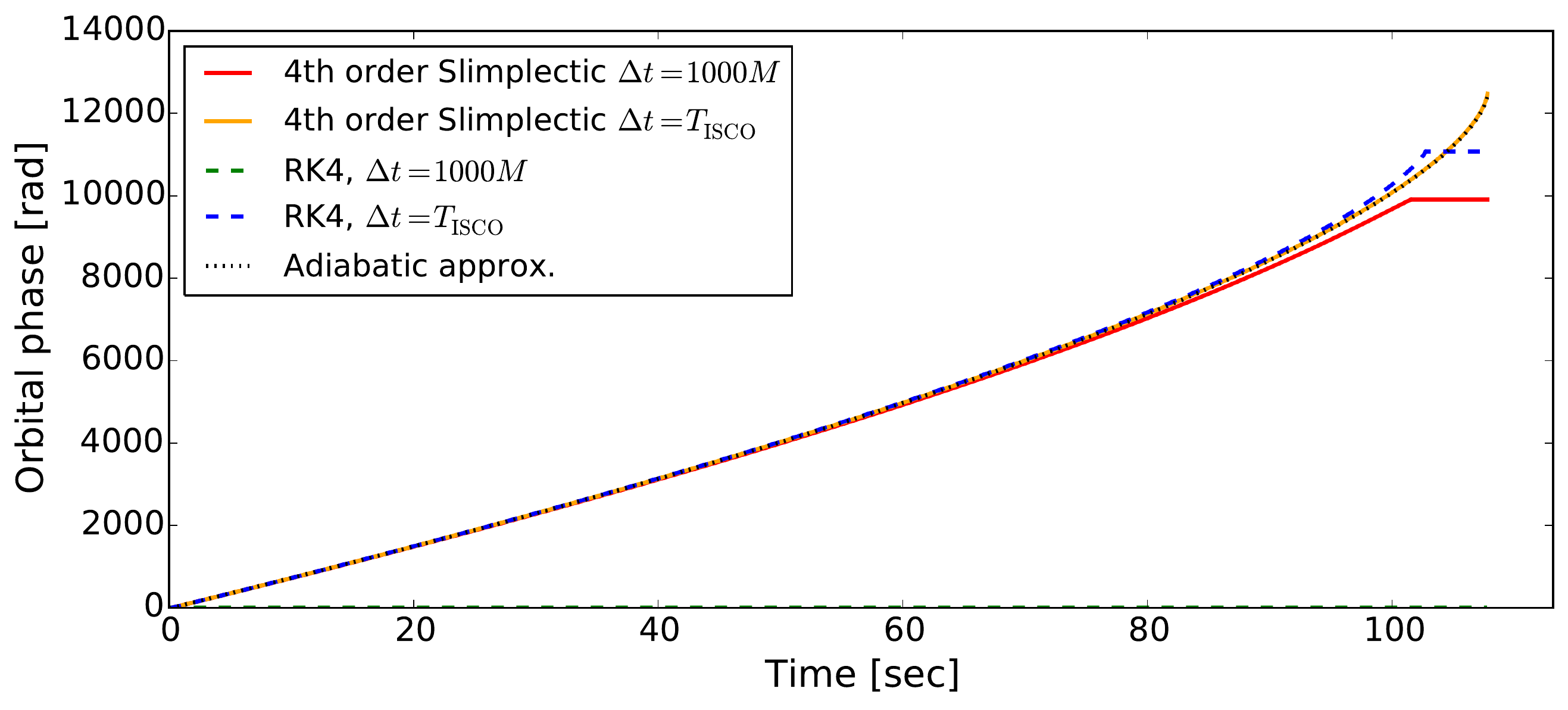}
\includegraphics[width=\columnwidth]{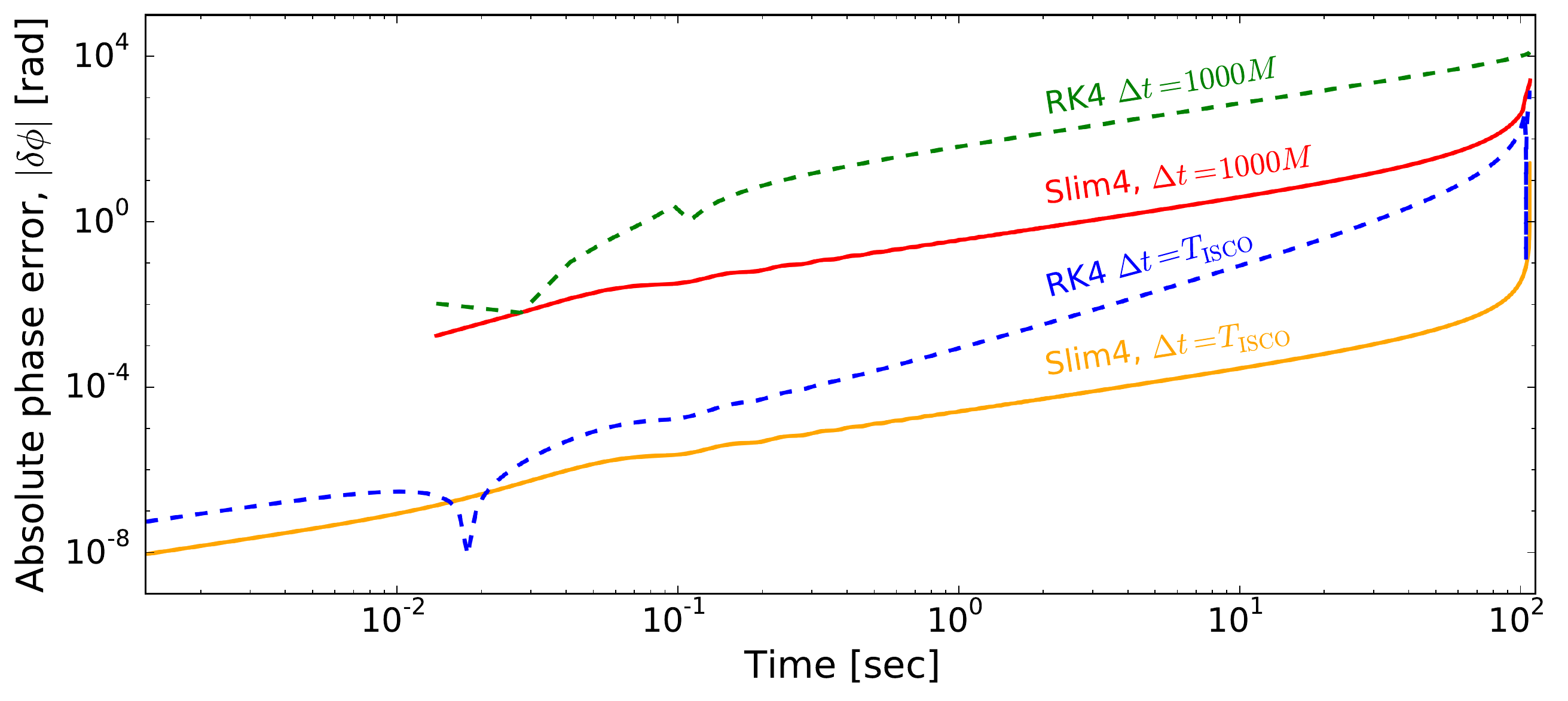}
\caption{%
{\bf Top:} The phase evolution of the integrators for the PN radiation reaction depicted in Figure \ref{fig:PNr}.  \\
{\bf Bottom:} Absolute orbital-phase errors. In this example, both slimplectic integrators (phase error $\propto t$), track the orbital phase much better
than the equivalent RK integrators (phase error $\propto t^2$). \label{fig:PNphi}
}
\end{figure}

In Figures~\ref{fig:PNr} and \ref{fig:PNphi} we examine the 4th-order RK and slimplectic
integrators for different choices of time steps, $\Delta t = 1000M$
and $\Delta t = T_{\rm ISCO}\simeq 92M$, the orbital period at the innermost
stable circular orbit. The initial orbital separation is $r_0=100M \approx 414$km and
corresponds to an orbital frequency of $\approx 11.5$Hz. 
We compare our numerical results to analytic solutions to our
toy PN equations in the adiabatic regime,
\begin{align}
	r_{\rm ad}(t) \!=\! \left( r_0 - \frac{ 256  }{ 5 } \nu M^3 t \right)^{\!1/4} {\hskip-0.1in},~ 
	\phi_{\rm ad}(t)  \!=\! \frac{ r_0^{5/2} - r_{\rm ad}^{5/2}(t)  }{32 \nu M^{5/2} }  \,,
	\nonumber
\end{align}
where the orbital period is assumed to be much smaller than
the radiation reaction time scale for the inspiral~\citep{Blanchet2014}.

All integrators begin to fail when the orbital period becomes 
roughly comparable to the time-step, although the slimplectic integrators 
can get significantly closer to this limit than the RK method (the RK integrator with 
time-step $\Delta t = 1000M$ immediately becomes unstable). In our
followup paper, we will demonstrate an adaptive time-stepping 
scheme that will allow efficient slimplectic integrations that also 
precisely evolve the energy.

\section{Discussion}

We have developed a new method of numerical integration that combines
the nonconservative action principle of~\citet{Galley2013, Galley2014}
with the variational-integrator approach of~\citet{Marsden2001}. 
These ``slimplectic'' integrators allow nonconservative effects to be
included in the numerical evolution, while still possessing the major
benefits of normally conservative symplectic integrators, particularly
the accurate long-term evolution of momenta and energy.  

The discrete equations of motion are found by varying a discretized nonconservative action and implicitly define the
slimplectic mapping $(q_n, \pi_n) \rightarrow (q_{n+1}, \pi_{n+1})$. Different choices of 
discretization generate different variational integrators.  Here we have focused
on implementing the Galerkin-Gauss-Lobatto (GGL) discretization, and demonstrating
its long-term accuracy using the damped harmonic oscillator,
Poynting-Robertson drag on a small particle, and a gravitational
radiation-reaction toy problem.
Our results also explain why the modification of the 2nd-order ``kick-drift-kick''
ansatz to include dissipative forces performs so accurately, as this
is equivalent to the lowest order version of the slimplectic GGL
method.

We have developed a demonstration \texttt{python} code,
\texttt{slimplectic},\textsuperscript{\ref{fn:repo}} which
generates slimplectic integrators for
arbitrary Lagrangians and nonconservative potentials.  Readers are
encouraged to test different physical systems of interest using this
publicly available code, but to separately implement problem specific
optimizations, particularly when solving the implicit equations of
motion.


{\it Acknowledgments }--- We thank A.~Cumming, A.~Archibald,  
D.~Tamayo, D.P.~Hamilton, M.C.~Miller, 
and the referee, W. Farr, for useful discussion.
D.T.~was supported by the Lorne Trottier Chair in Astrophysics and
Cosmology and CRAQ. 
C.R.G.~was supported in part by NSF grants CAREER
PHY-0956189 and PHY-1404569
at Caltech. L.C.S.~was supported by NASA through Einstein
Postdoctoral Fellowship Award Number PF2-130101.


\end{document}